\documentclass[fleqn,10pt]{wlscirep}
\title{Magnetic Properties Controlled by Interstitial or Interlayer Cations in Iron Chalcogenides}

\author[1]{Shan-Chang Tang}
\author[1,2]{Ming-Cui Ding}
\author[1,3,*]{Yu-Zhong Zhang}
\affil[1]{Shanghai Key Laboratory of Special Artificial Microstructure Materials and Technology,
School of Physics Science and engineering, Tongji University, Shanghai 200092, P.R. China}
\affil[2]{School of Physics and Optoelectronic Engineering, Ludong University,  Yantai 264025 P.R. China}
\affil[3]{Beijing Computational Science Research Center, Beijing 100084, China}

\affil[*]{yzzhang@tongji.edu.cn}

\begin{abstract}
By applying density functional theory calculations to iron chalcogenides, we find that magnetic order in Fe$_{1+y}$Te and magnetic instability at $(\pi,\pi)$ in K$_y$Fe$_2$Se$_2$ are controlled by interstitial and interlayer cations, respectively. While in Fe$_{1+y}$Te, magnetic phase transitions occur among collinear, exotic bicollinear and plaquette-ordered antiferronmagnetic states when the height of interstitial irons measured from iron plane or the concentration of interstitial irons is varied, the magnetic instability at $(\pi,\pi)$ which is believed to be responsible for the Cooper pairing in iron pnictides is significantly enhanced when $y$ is much smaller than $1$ in K$_y$Fe$_2$Se$_2$. Our results indicate that, similar to iron pnictides, itinerant electrons play important roles in iron chalcogenides, even though the fluctuating local moments become larger.
\end{abstract}
\begin{document}

\flushbottom
\maketitle
\thispagestyle{empty}

\section*{Introduction}

Though the research field of high-T$_c$ iron-based superconductors grows rapidly, the question of whether itinerant electrons or local moments are responsible for superconductivity and magnetism remains unsolved~\cite{Dai_Hu_Dagotto,Mazin_Singh,Singh_Du,FangWangEPL,Knolle_Eremin,Zhang_Opahle,Kang_Wang,Schmidt_Siahatgar,Fang_Xu,Ma_Ji,Si_Abrahams,Yin}. Recently, relevance of itinerant electrons to the physical properties of iron-based superconductors becomes questionable due to the discovery of iron chalcogenides. Specifically, in Fe$_{1+y}$Te, while the Fermi surfaces show nesting vector of $(\pi,\pi)$ from angular resolved photoemission spectroscopic study~\cite{Xia}, which supports a collinear antiferromagnetic (CAF) order at low temperature, neutron scattering analysis indicates a bicollinear antiferromagnetic (BAF) order~\cite{Bao,Sli}, which requires the Fermi surfaces nested at $(\pi,0)$. This contradiction casts doubt on the applicability of itinerant scenario of magnetism to iron chalcogenides. The itinerant scenario has been further challenged since high-T$_c$ superconductor K$_{0.8}$Fe$_2$Se$_2$ was synthesized~\cite{guochen}. There is no hole pocket at $\Gamma$ point~\cite{Zhang-Feng}, indicating that the itinerant scenario of superconductivity arising from nesting of the Fermi surfaces around $\Gamma$ and $M$ points may not be valid for iron chalcogenides. In contrast, it has been gradually accepted that itinerant electrons can not be ignored in high-T$_c$ iron pnictides~\cite{Zhao-Adroja-Dai,Wang-Li}. Therefore, a fundamental problem arises: are itinerant electrons irrelevant to the magnetism and superconductivity in iron chalcogenides, which is entirely different from the situation in iron pnictides?

In order to solve the problem, effects of interstitial Fe in Fe$_{1+y}$Te$_{1-x}$Se$_x$ and dopant A in the interlayer of A$_y$Fe$_2$Se$_2$ (A=K, Rb) were intensively investigated experimentally. It was found that, while varying small amounts of interstitial Fe in Fe$_{1+y}$Te$_{1-x}$Se$_x$ can tune the superconducting and magnetic properties~\cite{Bendele2010,Stock2012}, the actual composition of  superconducting fraction in A$_y$Fe$_2$Se$_2$ (A=K, Rb) is of $y=0.3(1)$ when A=Rb~\cite{Texier} and of $y=0.53$ when A=K~\cite{Carr-Dai}. However, the underlying physics of interstitial and interlayer cations in iron chalcogenides is still missing.

In fact, there was one leading work about the effect of interstitial Fe on the magnetic instability in the high-temperature paramagnetic phase of Fe$_{1+y}$Te~\cite{HanSavrasov}. It was proposed that the interstitial Fe provides additional electrons to Fe-Te layers, causing a raise of the Fermi level and as a consequence inducing a strong magnetic instability at $(\pi,0)$, which supports the itinerant origin of BAF order. Although there were strong debates on the amounts of doped electrons supplied by the interstitial Fe~\cite{PPSingh,HanSavrasovReply}, the proposal was confirmed after orbital degrees of freedom are taken into account~\cite{Ding-Zhang}. However, in the above researches, apart from the existing debates, simple rigid band approximation is employed when the effect of interstitial Fe is considered, lessening the reliability of the conclusions. Moreover, it remains unknown about the actual effect of interstitial Fe on the low-temperature antiferromagnetic state of Fe$_{1+y}$Te~\cite{ZhangSingh}.

\begin{figure}[ht]
\centering
\includegraphics[width=0.7\linewidth]{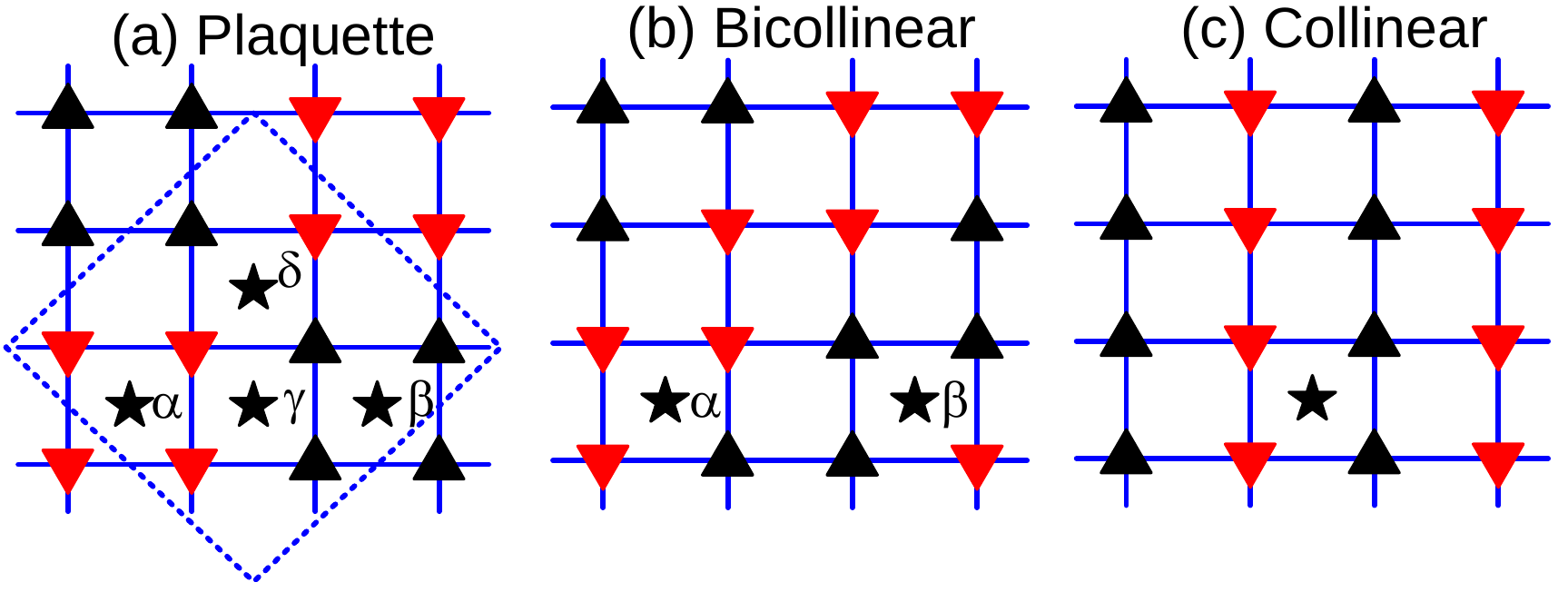}
\caption{Cartoons for different magnetically ordered states with interstitial Fe sitting in various inequivalent positions. Please note, those are the top view of Fe$_{1+y}$Te without Te atoms. The interstitial Fe denoted by star is out of the Fe plane denoted by triangles. Different colors represent for different spin directions. (a) plaquette antiferromagnetic state (PAF) with four inequivalent positions of interstitial Fe denoted by $\alpha$, $\beta$, $\gamma$, and $\delta$. The dashed line denotes a supercell of Fe$_9$Te$_8$, namely Fe$_{1.125}$Te. (b) bicollinear antiferromagnetic state (BAF) with two inequivalent positions of interstitial Fe denoted by $\alpha$, $\beta$. (c) collinear antiferromagnetic state (CAF).}
\label{Fig:one}
\end{figure}

Furthermore, a recent experimental study pointed out that the low-temperature phase of Fe$_{1+y}$Te could be of plaquette antiferromagnetic (PAF) order~\cite{Igor}, rather than the prevailing BAF one. Such a new proposal was supported by a theoretical analysis based on a Heisenberg J$_1$-J$_2$-J$_3$ model where quantum fluctuations were appropriately involved~\cite{Ducatman}. Then, another important question appears: which on earth is the ground state of Fe$_{1+y}$Te at low temperature, either the BAF or PAF order?

In this paper, we investigate the effect of interstitial and interlayer cations in iron chalcogenides by density functional theory (DFT) calculations. It is found that  magnetic ground state of Fe$_{1+y}$Te is determined not only by the concentration of interstitial Fe but also by its height away from Fe plane. While less amount of interstitial Fe favors the BAF state, more interstitial Fe leads to the PAF phase. Moreover, we for the first time discover that magnetic phase transitions among the BAF, PAF, and CAF states can be realized by tuning the height of interstitial Fe, indicating that the ground state property of Fe$_{1+y}$Te is susceptible to annealing process since the interstitial Fe may be quenched at different heights. Finally, we find that the magnetic instability at $(\pi,\pi)$ which is responsible for the Cooper pairs in iron pnictides is significantly enhanced when the concentration of interlayer alkali metal is reduced in A$_y$Fe$_2$Se$_2$. Our results indicate that, simliar to iron pnictides, itinerant electrons are also closely relevant to the magnetism and superconductivity in iron chalcogenides as long as correct chemical compositions are used.

\section*{Results}

First, we will study the effect of interstitial Fe in Fe$_{1+y}$Te. Three different magnetically ordered states, such as the PAF, BAF, and CAF states, are taken into account which are schematically depicted in Fig.~\ref{Fig:one}. The N\'{e}el antiferromagnetic ordered state and the ferromagnetic state were also investigated but were not shown here since the total energies are much higher than those of the three states we considered. Since the magnetically ordered states are taken into account, several inequivalent positions exist when putting the interstitial Fe out of Fe plane as shown in Fig.~\ref{Fig:one}. Throughout the paper, we use combination of the position of interstitial Fe and the magnetic pattern to uniquely define the state we studied. For example, $\delta$-PAF denotes the PAF state with interstitial Fe located in $\delta$ position.

\begin{figure}[ht]
\centering
\includegraphics[width=0.5\linewidth]{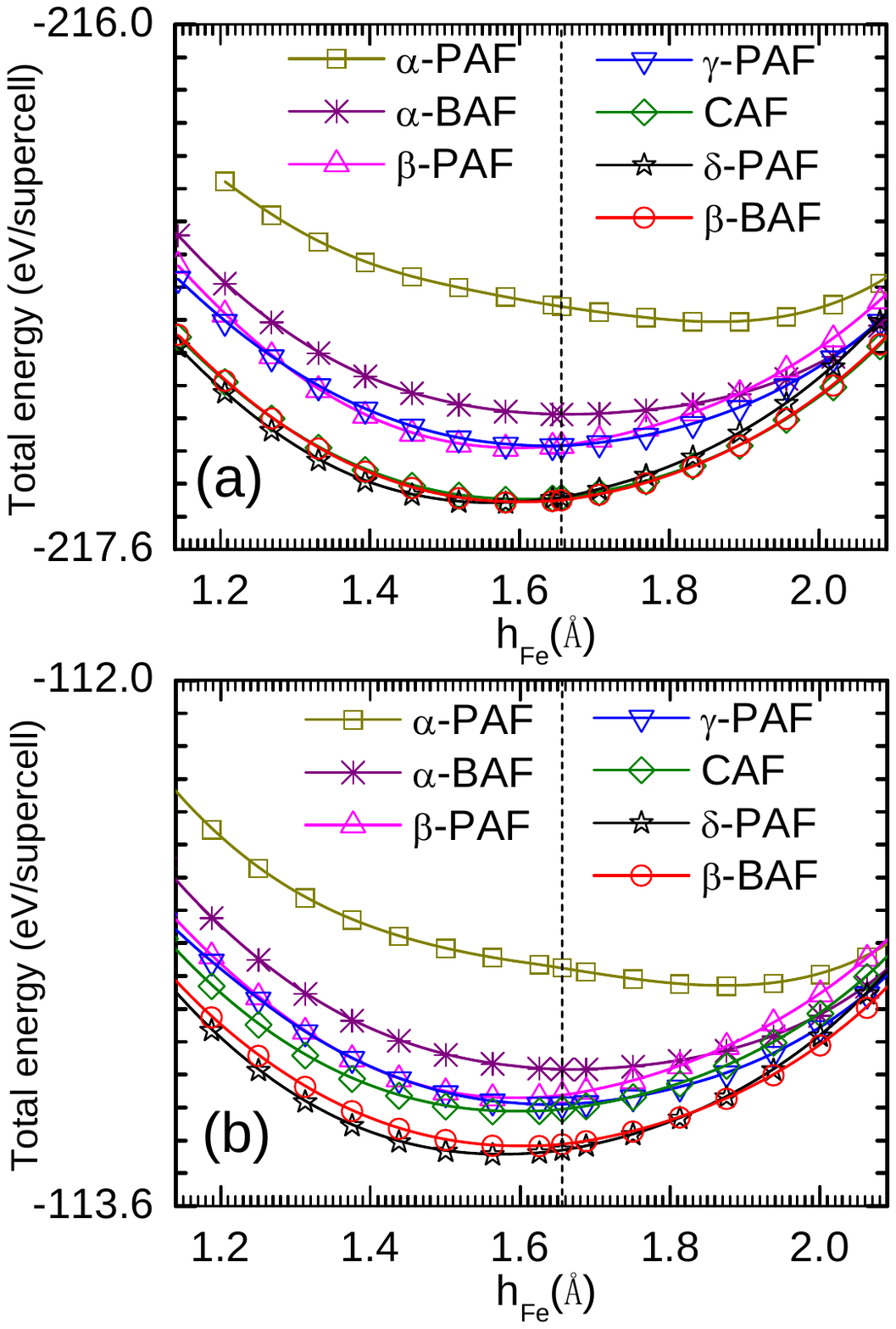}
\caption{Total energies of different magnetically ordered states shown in Fig.~\ref{Fig:one} as a function of height of interstitial Fe measured from Fe plane in the supercells of Fe$_{17}$Te$_{16}$ (a) and Fe$_9$Te$_8$ (b), which corresponds to a stoichiometry of Fe$_{1.0625}$Te and Fe$_{1.125}$Te, respectively.  The experimental value of the position of interstitial Fe~\cite{Sli} is marked by dashed line. }
\label{Fig:two}
\end{figure}

\begin{figure}[ht]
\centering
\includegraphics[width=0.5\linewidth]{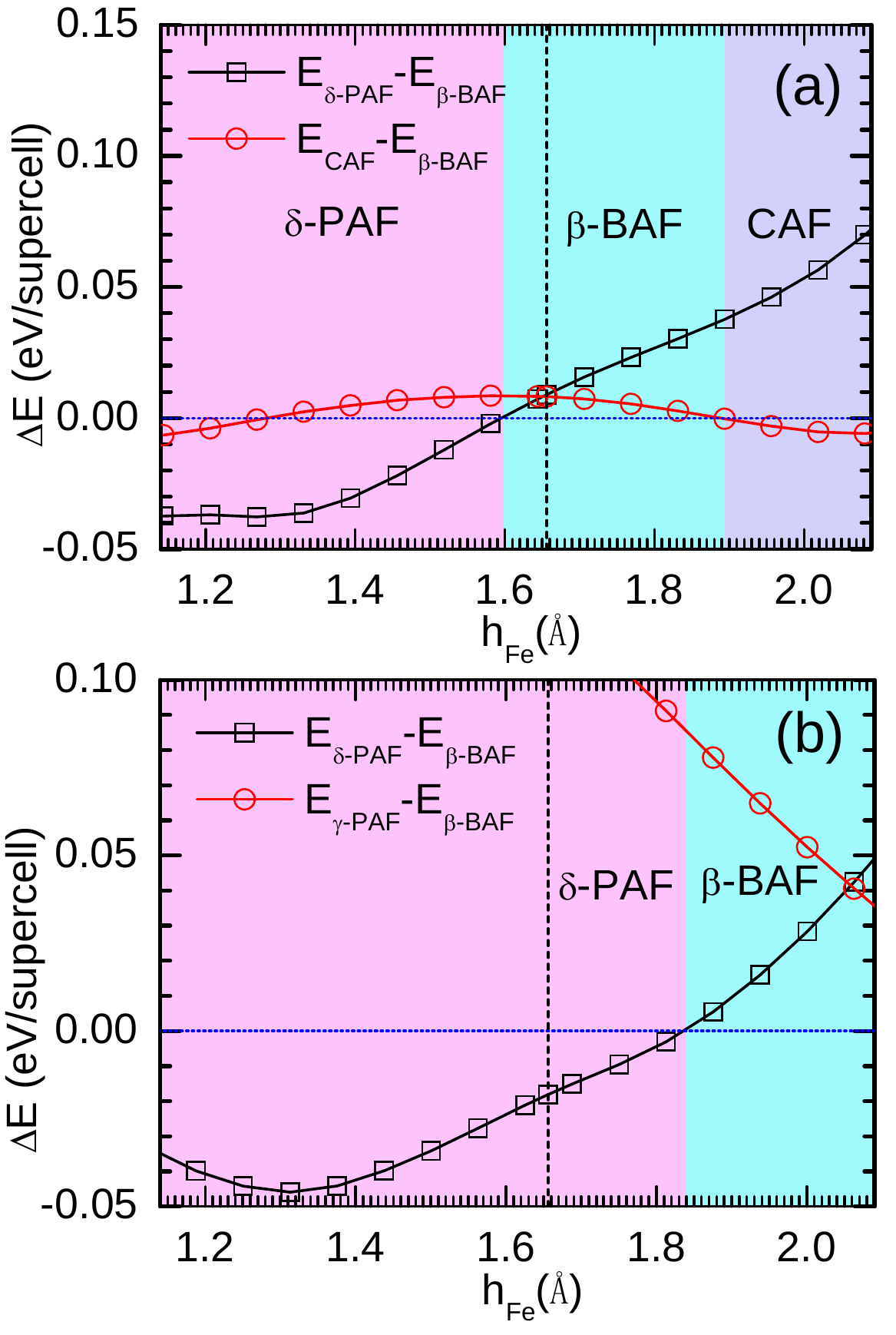}
\caption{Differences of the total energies among three competing phases in the case of Fe$_{17}$Te$_{16}$ (a) and Fe$_9$Te$_8$ (b) which corresponds to a stoichiometry of Fe$_{1.0625}$Te and Fe$_{1.125}$Te, respectively. The dashed line indicates the experimental value of the position of interstitial Fe~\cite{Sli}. Different phases are marked by different colors.}
\label{Fig:three}
\end{figure}

Fig.~\ref{Fig:two} shows total energies of the different states we considered as a function of the height of interstitial Fe measured from Fe plane. In the case of Fe$_{1.0625}$Te, as presented in Fig.~\ref{Fig:two} (a), there are three states competing with each other when the height of interstitial Fe is tuned around the experimental value of $h_{Fe}=1.6559 \AA$. Those are the CAF, $\beta$-BAF, and $\delta$-PAF states. In Fig.~\ref{Fig:three} (a), we present energy differences among these three states. It is found that the ground state is $\beta$-BAF when the experimental position of interstitial Fe is used, consistent with previous experimental observation~\cite{Sli}. While slightly tuning down the height of interstitial Fe will induce a phase transition from the $\beta$-BAF state to the $\delta$-PAF state, larger distance between interstitial Fe and Fe plane will favor the CAF state.

The situation is quite different when the concentration of interstitial Fe is increased. As is shown in Fig.~\ref{Fig:two} (b), only two states, i.e., the $\beta$-BAF and $\delta$-PAF states, compete with each other. From Fig.~\ref{Fig:three} (b), it is found that the ground state at $h_{Fe}=1.6559 \AA$ is now located deeply inside the $\delta$-PAF phase, which strongly suggests that the ground state of Fe$_{1.125}$Te is most probably of the $\delta$-PAF state, although the experimental height of interstitial Fe measured from Fe plane in the PAF state is currently inaccessible experimentally.

\begin{figure}[ht]
\centering
\includegraphics[width=0.5\linewidth]{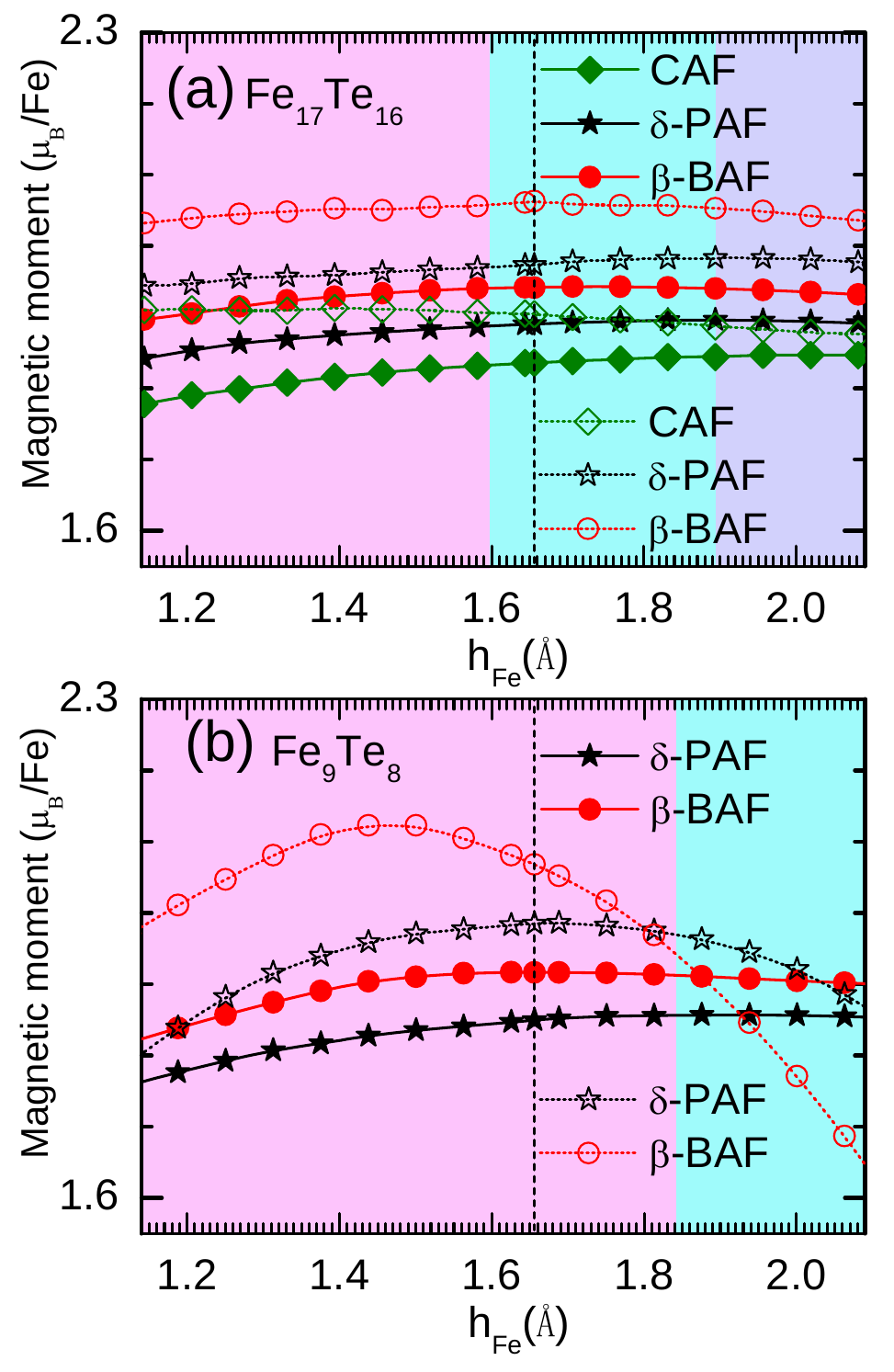}
\caption{Magnetic moments of the competing magnetic states in the case of Fe$_{17}$Te$_{16}$ (a) and Fe$_9$Te$_8$ (b) which corresponds to a stoichiometry of Fe$_{1.0625}$Te and Fe$_{1.125}$Te, respectively. Solid symbols represent the magnetic moment per iron in the iron plane while empty symbols denotes the magnetic moment on the interstitial iron. The dashed line indicates the experimental value of the position of interstitial Fe~\cite{Sli}. Different phases are marked by different colors as indicated in Fig.~\ref{Fig:three}.}
\label{Fig:fournew}
\end{figure}

Fig.~\ref{Fig:fournew} presents the calculated magnetic moments of the competing magnetic states for above two concentrations. It is found that the magnetic moments are more strongly dependent on the height of the interstitial Fe in Fe$_{1.125}$Te than in Fe$_{1.0625}$Te. And for both concentrations, the magnetic moments are quite different among different competing phases. The calculated magnetic moments are in good agreement with the experimental results when the experimental position of the interstitial Fe is used~\cite{Bao,Sli,Rodriguez,Iikubo}.

\begin{table}[ht]
\centering
\begin{tabular}{|l|l|l|}
\hline
$h_{Fe} (\AA)$ & $\Delta E^{\beta-BAF}_{Kin(Int)}$ & $\Delta E^{\delta-PAF}_{Kin(Int)}$ \\
\hline
2.0815   &  15.002(-14.996)                  &  10.490(-10.414)  \\
\hline
1.6559   &  17.109(-17.117)                  &  10.502(-10.502)  \\
\hline
1.5188   &  17.183(-17.191)                  &   9.995(-10.015)   \\
\hline
\end{tabular}
\caption{\label{Tab:one}Kinetic (interaction) energies of $\beta$-BAF and $\delta$-PAF states with respect to that of CAF state at three different heights of interstitial Fe, in eV/supercell, defined as $\Delta E^{\phi}_{Kin(Int)}(h_{Fe})=E^{\phi}_{Kin(Int)}(h_{Fe})-E^{CAF}_{Kin(Int)}(h_{Fe})$.}
\end{table}

While above investigations lead to an important conclusion that the interstitial Fe controls the magnetic properties of Fe$_{1+y}$Te, it is still unclear if the above mentioned phase transitions are only induced by minimization of interaction energy. To clarify this issue, we separate the kinetic energy ($E_{Kin}$) from the interaction energy ($E_{Int}$) in Fe$_{1.0625}$Te. The differences of kinetic (interaction) energies between the $\beta$-BAF and CAF states, as well as the $\delta$-PAF and CAF states, are shown in Table~\ref{Tab:one} at three different heights of $h_{Fe}=2.0815$, $1.6559$, and $1.5188 \AA$. It is found that competition among these three phases is a result of balance between gain and loss of kinetic and interaction energies, rather than minimization of interaction energy alone. For example, at $h_{Fe}=2.0815 \AA$, although choosing the $\beta$-BAF or $\delta$-PAF states can both save interaction energy compared to the CAF state, it will lose more kinetic energy. Therefore, in this case, minimization of kinetic energy plays a dominant role in the appearance of CAF state. At the experimental value of $h_{Fe}=1.6559 \AA$, reduction of interaction energy becomes dominant and the $\beta$-BAF state becomes favorable. At $h_{Fe}=1.5188 \AA$, decrease of interaction energy is more than increase of kinetic energy in both the $\beta$-BAF and $\delta$-PAF states. Further comparing the energies of the $\beta$-BAF to the corresponding ones of the $\delta$-PAF states, we find that choosing the $\delta$-PAF state will gain kinetic energy of $7.188$~eV while lose interaction energy of $7.176$~eV. As a consequence, the $\delta$-PAF state appears, due to the subtle balance of minimizing kinetic energy and interaction energy simultaneously. Our results point to a fact that kinetic energy of the itinerant electrons plays an important role in determining the magnetic properties of Fe$_{1+y}$Te, rather than interaction energy alone.

\begin{figure}[ht]
\centering
\includegraphics[width=0.5\linewidth]{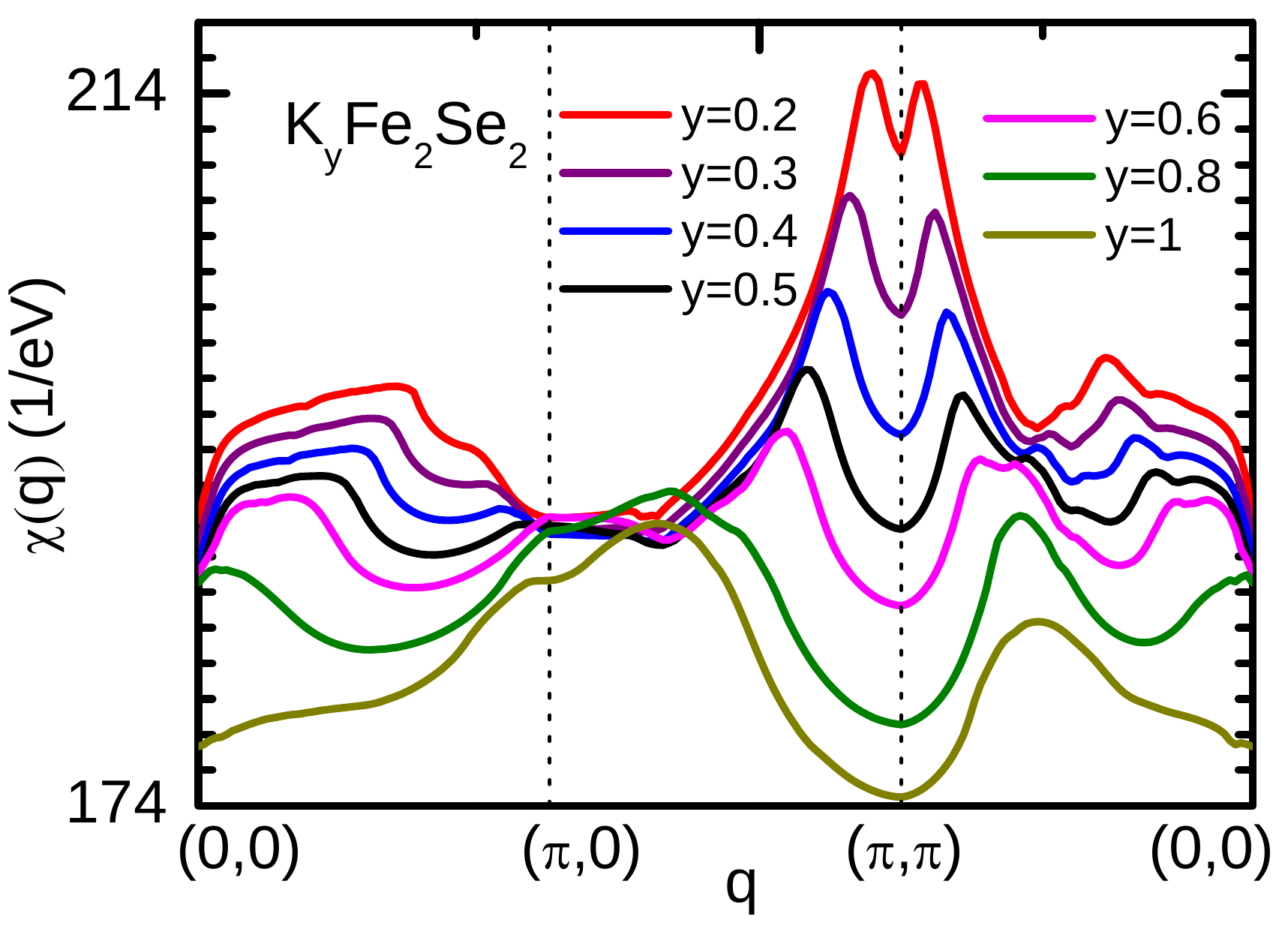}
\caption{Evolution of the momentum dependent Pauli susceptibilities as a function of potassium concentration in K$_y$Fe$_2$Se$_2$. Starting from $y>0.6$, a prominent peak around $(\pi,\pi)$ develops, indicating a strong magnetic instability with wave vector of $(\pi,\pi)$ appears, which supports the itinerant scenario of superconductivity in K$_y$Fe$_2$Se$_2$.}
\label{Fig:four}
\end{figure}

Then, we will study the effect of interlayer cations in A$_y$Fe$_2$Se$_2$ with A=Rb and K. According to the experimental findings that the actual composition of the superconducting phase is Rb$_{0.31}$Fe$_2$Se$_2$ and K$_{0.53}$Fe$_2$Se$_2$~\cite{Carr-Dai,Texier}, rather than the nominal one where the content of alkali metals $y$ is much large, we will study the effect of different concentration of alkali metal on the Pauli susceptibility at $(\pi,\pi)$ which is a direct evidence for possible appearance of superconductivity in iron-based superconductors~\cite{Mazin_Singh,FangWangEPL,Zhang_Opahle}. Here, the concentration $y$ is tuned from $0.2$ to $1$ by shifting the Fermi level. Fig.~\ref{Fig:four} presents the evolution of momentum dependent Pauli susceptibilities as a function of $y$ in K$_y$Fe$_2$Se$_2$. It is found that at $y=0.8$ and $1$, the susceptibility exhibit a global minimum at $(\pi,\pi)$, indicating that there is no magnetic instability around $(\pi,\pi)$, which is consistent with the lack of hole pocket at $\Gamma$ point observed experimentally~\cite{Zhang-Feng}. Starting from $y>0.6$, a strong peak develops around $(\pi,\pi)$. Further decreasing the concentration of potassium atoms down to $y=0.2$, the peak gets stronger and its position is even closer to $(\pi,\pi)$. This indicates the appearance of a strong magnetic instability around $(\pi,\pi)$ at small $y$, which supports the itinerant scenario of superconductivity, reminiscent of similar situations in other iron-based superconductors~\cite{Mazin_Singh,FangWangEPL,Zhang_Opahle}.

\section*{Discussion}

Finally, we will discuss the relevance of our results to the experiments. As we mentioned above, the prevailing BAF order observed in Fe$_{1.068}$Te~\cite{Sli} and Fe$_{1.076}$Te~\cite{Bao} was questioned after inelastic neutron scattering study on Fe$_{1.1}$Te was performed~\cite{Igor}, which indicates a new PAF order. The contradiction can be simply resolved, as we find that the magnetic property in Fe$_{1+y}$Te is controlled by the concentration of interstitial Fe. Smaller amount of interstitial Fe favors the BAF order while larger supports the PAF order. Our results can also be applied to explain the experimental observations that superconducting properties of Fe$_{1+y}$Te$_{1-x}$Se$_x$ can be tuned by the Fe content~\cite{Bendele2010,Stock2012}. This is due to the fact that the interstitial Fe always tends to favor either the BAF or the PAF states but suppresses the CAF state which is the parent state for superconductivity. Moreover, our results indicate that the magnetism or superconductivity in Fe$_{1+y}$Te$_{1-x}$Se$_x$ may also be affected by the annealing process when samples are prepared since the interstitial Fe may be quenched at different heights, leading to different type of magnetic instabilities. Please note that our results are obtained based on DFT calculations where fluctuating local moments are not involved but can well account for puzzling experiments, indicating that the itinerant electrons are important to the magnetism in Fe$_{1+y}$Te.

In K$_y$Fe$_{2-x}$Se$_2$, phase separation has already been observed experimentally~\cite{XueNP2012}. However, no theoretical investigation was performed on the actual composition of the superconducting phase. We for the first time theoretically discovered that strong magnetic instability also appears around $(\pi,\pi)$ when the genuine superconducting compound is taken into account, similar to what happens in iron pnictides. This indicates that iron chalcogenides and iron pnictides share the same scenario of superconductivity where itinerant electrons are important.

In conclusion, we study the effects of interstitial and interlayer cations on the magnetic properties of iron chalcogenides by DFT calculations. We find that the magnetically ordered state in Fe$_{1+y}$Te and the magnetic instability in K$_y$Fe$_2$Se$_2$ are controlled by the interstitial Fe and the interlayer dopant K, respectively. Our results strongly indicate that the scenario for magnetism and superconductivity should be the same for both iron pnictides and iron chalcogenides, where itinerant electrons can not be ignored, rather than the Mottness alone. Moreover, our work has a wide implication that nominal composition may not be a good starting point for understanding real materials. Attention should also be paid to other interesting systems, like the potassium doped picene~\cite{Mitsuhashi,Yan2014}. Our work opens a new perspective in understanding the magnetism and superconductivity of iron-based superconductors. Further studies can be done to take into account the interactions between interstitial Fe in Fe$_{1+y}$Te$_{1-x}$Se$_x$ or to establish a general theory for both iron pnictides and iron chalcogenides based on the fact that all the bulk iron-based superconductors can be described within weak coupling limit.

\section*{Methods}

Our results on Fe$_{1+y}$Te are obtained based on DFT calculations with local spin density approximation (LSDA) and the project augmented wave method~\cite{Blochl, Kresse-Joubert} as implemented in the VASP code~\cite{Kresse-Furthm1, Kresse-Furthm2}. A kinetic-energy cutoff of $508$~eV was used to obtain converged energy within $0.01$~meV. To simulate the partially occupied interstitial Fe, we used $2\sqrt{2}\times2\sqrt{2}$ and $2\times2$ supercells of $\alpha$-FeTe (two formulas per cell) with one interstitial Fe atom placed at $(0.25,0.25,z_{Fe})$ site. This corresponds to a stoichiometry of Fe$_{1.0625}$Te and Fe$_{1.125}$Te, respectively. The experimental lattice parameters of $a=b=3.81234(8) \AA$, $c=6.2517(2) \AA$, $z_{Te}=0.2829(4)$~\cite{Sli} were used in our calculations. The internal position $z_{Fe}$ of interstitial Fe will be tuned around the experimental value~\cite{Sli}. The height of interstitial Fe ($h_{Fe}$) measured from Fe plane can be obtained by $h_{Fe}=(1-z_{Fe})c$. A $9\times9\times9$ grid was used for the k-point sampling of the Brillouin zone. The $3s3p3d4s$ states in Fe and the $5s5p$ states in Te are treated as valence states. Part of our results were double-checked by the full potential linearized augmented plane-wave method as implemented in the WIEN2k code~\cite{balaha}. Very good agreement was found between these two methods.

The reason why we use LSDA rather than spin polarized generalized gradient approximation (GGA) for Fe$_{1+y}$Te is that the static magnetic moment calculated within LSDA agrees well with the experimental results~\cite{Bao,Sli,Rodriguez,Iikubo} while GGA always overestimates it. As we are interested in the magnetic phase transitions where the static magnetic moment is the dominant quantity in our study, we use LSDA for all the calculations of Fe$_{1+y}$Te. Please note, although large local moment is observed experimentally, it is fluctuating dynamically and does not contribute to the static long-range magnetic order. moreover, it has already been suggested that LSDA is better than GGA when calculating magnetic ground state in iron-pnictides~\cite{MazinJohannes,OpahleKandpal}.

Calculations of the $\mathbf{q}$-dependent Pauli susceptibility at $\omega$=0 within the
constant matrix element approximation~\cite{Mazin_Singh,FangWangEPL,Zhang_Opahle} in K$_y$Fe$_2$Se$_2$ were done by the WIEN2k code with experimental structure of $a=b=3.9136(1) \AA$, $c=14.0367(7) \AA$, $z_{Se}=0.3539(2)$~\cite{guochen}. The $\mathbf{q}$-dependent Pauli susceptibility at $\omega$=0 within the constant matrix element approximation is defined as
\begin{equation*}
\chi _0\left( \mathbf{q}\right) =-\sum\limits_{\mathbf{k}\alpha \beta }\frac{%
f\left( \varepsilon _{\mathbf{k}\alpha }\right) -f\left( \varepsilon _{%
\mathbf{k+q}\beta }\right) }{\varepsilon _{\mathbf{k}\alpha }-\varepsilon _{%
\mathbf{k+q}\beta }+i\delta }  \label{Paulisusceptibility}
\end{equation*}
where $\alpha$ and $\beta$ are band indexes and $q$ and $k$ are momentum vectors in the Brillouin zone. Here $f(E)$ is the Fermi distribution function and $\varepsilon _{\mathbf{k}\alpha}$ is the energy of band $\alpha$ at momentum vector of $k$. A three dimensional grid of $128\times128\times128$ $\mathbf{k}$ points in the full Brillouin zone is employed for calculation of the susceptibility. The Lorentzian broadening factor $\delta=0.01$ is used. Please note, in the calculation of the Pauli susceptibility, both LDA and GGA functionals give same results~\cite{Zhang_Opahle,Ding-Zhang,MgFeGe}.

The separation of kinetic energy from interaction energy is performed according to Eq.~(1), Eq.~(48), and Eq.~(49) of Ref. 37 where different parts of the total energy individually calculated by the VASP code are explicitly defined.

Finally, we should mention that throughout the paper, we only use the experimental structures without optimization. This is due to the fact that lattice optimization in iron-based superconductors always leads to remarkable deviations from experimental structure, which in turn strongly affects the electronic structure~\cite{MazinJohannes}. On the other hand, it has been noticed that the electronic structure observed experimentally can be well accounted for with the experimental lattice structure~\cite{Xia,HanSavrasov}, rather than the optimized one.

\section*{Acknowledgements}

We acknowledge Wei-Guo Yin for helpful discussions. This work is supported by National Natural Science Foundation of China (Nos. 11174219 and 11474217), Program for New Century Excellent Talents in University (NCET-13-0428), and the Program for Professor of Special Appointment (Eastern Scholar) at Shanghai Institutions of Higher Learning as well as the Scientific Research Foundation for the Returned Overseas Chinese Scholars, State Education Ministry. M.-C. also acknowledges supports from Shandong Provincial Natural Science Foundation, China (No. ZR2014AP014).

\end{document}